\documentclass[twocolumn,secnumarabic,amssymb,amsmath,nofootinbib,tightenlines,showpacs,floatfix,superscriptaddress,aps,prl]{revtex4-1}
\usepackage{amsfonts,amsbsy,graphicx,bm,color,txfonts,dcolumn,times}

\begin{document}
\title{Omnidirectionally Bending to the Normal in $\epsilon$-near-Zero Metamaterials}
\author{Simin Feng}
\email{simin.feng@navy.mil}
\affiliation{Michelson Lab, Physics Division, Naval Air Warfare Center, China Lake, California 93555}
\date{\today}

\begin{abstract}
Contrary to conventional wisdom that light bends away from the normal at the interface when it passes from high to low refractive index media, here we demonstrate an exotic phenomenon that the direction of electromagnetic power bends towards the normal when light is incident from arbitrary high refractive index medium to $\epsilon$-near-zero metamaterial.  Moreover, the direction of the transmitted beam is close to the normal for all angles of incidence.  In other words, the electromagnetic power coming from different directions in air or arbitrary high refractive index medium can be redirected to the direction almost parallel to the normal upon entering the $\epsilon$-near-zero metamaterial.  This phenomenon is counterintuitive to the behavior described by conventional Snell's law and resulted from the interplay between $\epsilon$-near-zero and material loss.  This property has potential applications in communications to increase acceptance angle and energy delivery without using optical lenses and mechanical gimbals.  \\

\end{abstract}


\maketitle
\begin{figure}[htb]
\centering\includegraphics[width=.4\textwidth]{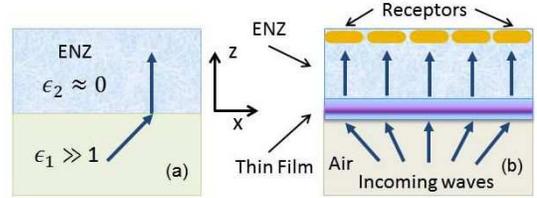}
\caption{(Color online) (a) A plane wave is incident from arbitrary high permittivity ($\epsilon_1\gg1$) medium to ENZ ($\epsilon_2\approx0$) metamaterial.  (b) Incoming waves in air from different directions all bend to the normal upon entering the ENZ medium.  A nanoplasmonic thin film is superimposed on the ENZ material to enhance the transmission.  Receptors or photocells are embedded in the ENZ metamaterial.}
\label{Fig1DTdraw}
\end{figure}

Bending of light towards the normal when it passes from low to high refractive index media is one of the fundamental phenomena in optics.  As a manifestation of this phenomenon, directive emission into air by a source inside the material with vanishingly small permittivity, known as $\epsilon$-near-zero (ENZ) metamaterials, has been demonstrated \cite{Enoch}.  With other intriguing properties, such as ultrathin waveguides \cite{Silveir,Liu,Edwards,Adams,Alu1}, diffraction-suppressed propagation and self-collimation \cite{Feng,Mocella1,Mocella2,Polles}, the ENZ materials have gained prominence as useful components to tailor antenna radiations \cite{Alu2,Halterman,Saenz,Mart}.  Previous studies on ENZ-directive emission have been focused on the radiation from low ($\epsilon\approx0$) to high (air) refractive index media \cite{Enoch,Ziolkow,Yuan,Jin,Lovat}, where the directive transmission can be intuitively understood from Snell's law that dictates the light bending towards the normal as it passes from low to high refractive index media.  
From the reciprocal theorem, for the radiation from high to low refractive index materials, the transmitted beam should spread out into grazing angles as the result of bending away from the normal.  Contrary to this conventional behavior, in this paper we will demonstrate an exotic phenomenon that electromagnetic (EM) power can bend towards the normal when light passes from arbitrary high ($\epsilon_1\gg1$) to low ($\epsilon_2\approx0$) refractive index media as shown in Fig.~\ref{Fig1DTdraw}a.  Furthermore, the direction of the transmitted beam is close to the normal for all angles of incidence.  More interestingly, this counterintuitive to conventional Snell's law behavior is induced by material loss.  The interplay between ENZ and loss leads to unusual wave interaction.  This phenomenon can be used to project EM power coming from different directions to one direction to the receivers as shown in Fig.~\ref{Fig1DTdraw}b, where the incoming waves all bend to the normal pointing to the receptors upon entering the ENZ medium.  
A plasmonic thin film is superimposed on the ENZ material to enhance the transmission through structural resonances.  For all the incoming directions including grazing angles, the transmitted powers impinge normally to the receptors or photocells embedded in the ENZ medium, effectively increasing the acceptance angle and energy transfer.

\begin{figure}[htb]
\centering\includegraphics[height=2.3in, width=3in]{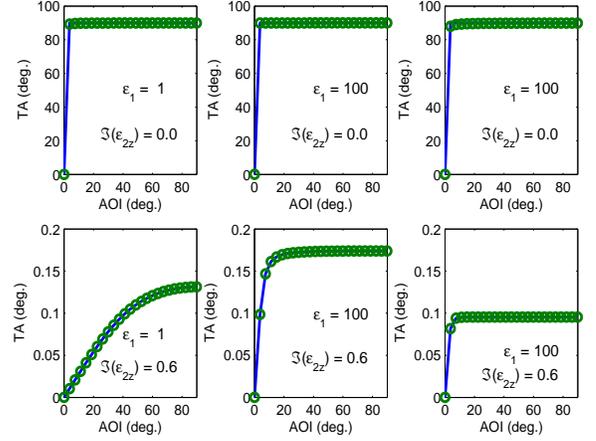}
\caption{(Color online) Transmission angle of the Poynting vector versus AOI.  Left and middle panels: anisotropic ENZ material with $\epsilon_{2x}=1$ and $\epsilon_{2z}=0.001+i\epsilon_{2z}^i$.  Right panels: isotropic ENZ material with $\epsilon_{2x}=\epsilon_{2z}=0.001+i\epsilon_{2z}^i$.  Top panels: $\epsilon_{2z}^i=0$.  Bottom panels: $\epsilon_{2z}^i=0.6$.  Left panels: $\epsilon_1=1$.  Middle and right panels: $\epsilon_1=100$.  A good agreement between the numerical (blue-solid) and analytical (green-circles) results.  The material loss switches the TA from the grazing angle $90^\circ$ (top panels) to the near-zero angle (bottom panels) for all the AOI.}
\label{Fig2DTang2}
\end{figure}

Our derivation is based on anisotropic media.  The results can be applied to isotropic materials.  Assuming a harmonic time dependence $\exp(-i\omega t)$ for the EM field, from Maxwell's equations, we have
\begin{eqnarray}
\label{Maxw}
\begin{split}
\nabla\times\bigl( \bar{\bar\mu}_n^{-1} \cdot \nabla\times{\bm E}\bigr)  &=\, k_0^2\bigl(\bar{\bar\epsilon}_n \cdot{\bm E}\bigr)  \,,\\
\nabla\times\bigl( \bar{\bar\epsilon}_n^{-1} \cdot \nabla\times{\bm H}\bigr)  &=\, k_0^2\bigl(\bar{\bar\mu}_n \cdot{\bm H}\bigr)  \,,
\end{split}
\end{eqnarray}
where $k_0=\omega/c$; and the $\bar{\bar\epsilon}_n$ and $\bar{\bar\mu}_n$ are, respectively, the permittivity and permeability tensors for each uniform region ($n=1,2,\cdots$), which in the principal coordinates can be described by
\begin{eqnarray}
\bar{\bar\epsilon}_n &=& \epsilon_{nx}\hat{\bm x}\hat{\bm x} + \epsilon_{ny}\hat{\bm y}\hat{\bm y} + \epsilon_{nz}\hat{\bm z}\hat{\bm z}  \,,\cr
\bar{\bar\mu}_n &=& \mu_{nx}\hat{\bm x}\hat{\bm x} + \mu_{ny}\hat{\bm y}\hat{\bm y} + \mu_{nz}\hat{\bm z}\hat{\bm z}  \,.
\end{eqnarray}
Consider transverse magnetic (TM) modes, corresponding to non-zero field components $H_y$, $E_x$, and $E_z$.  The magnetic field $H_y$ satisfies the following wave equation:
\begin{equation}
\frac{1}{\epsilon_z} \frac{\partial^2H_y}{\partial x^2} + \frac{1}{\epsilon_x} \frac{\partial^2H_y}{\partial z^2} + k_0^2\mu_yH_y = 0  \,,
\end{equation}
which permits solutions of the form $\psi(z)\exp(i\beta x)$.  Here the transverse wave number $\beta$ is determined by the incident wave, and is conserved across the interface of different regions,
\begin{equation}
\label{beta}
\beta^2 = k_0^2\epsilon_{nz}\mu_{ny} - \alpha_n^2 \frac{\epsilon_{nz}}{\epsilon_{nx}}  \,,\hskip.2in  (n=1,2,\cdots)  \,,
\end{equation}
where $\alpha_n$ is the wave number in the $z$ direction.  The functional form of $\psi(z)$ is either a simple exponential $\exp(i\alpha_nz)$ for the semi-infinite regions or a superposition of $\cos(\alpha_nz)$ and $\sin(\alpha_nz)$ terms for the bounded regions along the $z$ direction.  The other two components $E_x$ and $E_z$ can be solved from $H_y$ using Maxwell's equations.  By matching boundary conditions at the interfaces, i.e., the continuity of $H_y$ and $E_x$, the electromagnetic field can be derived in each region; and then the Poynting vector $\bm{S}$ can be computed from $\bm{S}=\Re(\bm{E}\times\bm{H}^*)$.  In anisotropic materials, the direction of the Poynting vector is different from that of the phase front of the field.  Here, only the direction of the Poynting vector is considered since it is associated with the energy transport.  The angle ($\theta_S$) of the Poynting vector is measured from the Poynting vector to the surface normal, and is given by $\theta_S=\tan^{-1}(S_x/S_z)$.  In Fig.~\ref{Fig1DTdraw}a, the input medium is isotropic material with permittivity $\epsilon_1$; the output medium is ENZ material ($\epsilon_2\approx0$).  In the following, both anisotropic ($\epsilon_{2x}\ne\epsilon_{2z}$) and isotropic ($\epsilon_{2x}=\epsilon_{2z}$) ENZ materials will be considered.  
Figure~\ref{Fig2DTang2} illustrates the effect of loss of the ENZ-materials on the transmission angle (TA), which is plotted against angle of incidence (AOI) with and without loss for the different permittivity ($\epsilon_1$) of the input medium.  In the top panels, when the loss is zero $\bigl(\Im(\epsilon_{2z})=0\bigr)$, the transmission angle is the grazing angle $90^\circ$ except for the normal incidence.  This behavior is complied with conventional Snell's law.  In the bottom panels, with a moderate loss $\Im(\epsilon_{2z})=0.6$, the transmission angle switches to near zero (normal direction) for all angles of incidence, leading to collimated transmission in the normal direction.  This switching phenomenon persists even for the much
higher permittivity ($\epsilon_1=100$) of the input medium (middle and right panels). 

To understand this loss-induced switching behavior, let's analyze the transmission angle ($\theta_S$), which is given by
\begin{equation}
\label{AngPoyn}
\tan(\theta_S)=\frac{S_x}{S_z} = \frac{\Re\left(\dfrac{\bar\beta}{\epsilon_{2z}}\right)} {\Re\left[\sqrt{\dfrac{\mu_{2y}}{\epsilon_{2x}} 
	- \dfrac{\bigl(\bar\beta\bigr)^2}{\epsilon_{2x}\epsilon_{2z}}} \right]}  \,,
\end{equation}
where $\bar\beta\equiv\beta/k_0$, and $\bar\beta$ (real) is determined by the incidence angle.  
The transmission angle of the Poynting vector depends only on the input and output media.  In the case of $\epsilon_{2x}\rightarrow0$ and $\epsilon_{2z}$ finite, Eq.~(\ref{AngPoyn}) indicates $\theta_S\rightarrow0^\circ$ (normal direction).  For the case of $\epsilon_{2z}\rightarrow0$ and $\epsilon_{2x}$ finite and the case of isotropic ENZ material with $\epsilon_{2x}=\epsilon_{2z}\rightarrow0$, the analysis is more involved.  The numerator of Eq.~(\ref{AngPoyn}) can be written as
\begin{equation}
\label{nume}
\Re\left(\frac{\bar\beta}{\epsilon_{2z}}\right) = \frac{\bar\beta\,\epsilon_{2z}^r}{|\epsilon_{2z}|^2}  \,,
\end{equation}
where $\epsilon_{2z}^r\equiv\Re(\epsilon_{2z})$.  Assuming $\mu_{2y}$ is real, the denominator of Eq.~(\ref{AngPoyn}) becomes
\begin{equation}
\label{deno}
\Re\left[\sqrt{\dfrac{\mu_{2y}}{\epsilon_{2x}} - \dfrac{\bigl(\bar\beta\bigr)^2}{\epsilon_{2x}\epsilon_{2z}}} \right] = \frac{a\,\bar\beta}{|\epsilon_{2x}\epsilon_{2z}|}  \,,
\end{equation}
where
\begin{equation}
a^2 = \frac{1}{2} \left(A\epsilon_{2x}^r + B|\epsilon_{2x}| - \epsilon_{2x}^r\epsilon_{2z}^r + \epsilon_{2x}^i\epsilon_{2z}^i\right)  \,,
\end{equation}
where $\epsilon_{2z}^i\equiv\Im(\epsilon_{2z})$, $\epsilon_{2x}^r\equiv\Re(\epsilon_{2x})$, $\epsilon_{2x}^i\equiv\Im(\epsilon_{2x})$, and
\begin{equation}
A\equiv \frac{|\epsilon_{2z}|^2\mu_{2y}}{\bigl(\bar\beta\bigr)^2}  \,,\hspace{.12in}  B = \sqrt{|\epsilon_{2z}|^2 - 2A\epsilon_{2z}^r + A^2}  \,.
\end{equation}
Thus, the transmission angle ($\theta_S$) becomes
\begin{equation}
\label{tantheS}
\tan(\theta_S) = \frac{|\epsilon_{2x}|\epsilon_{2z}^r}{a\,|\epsilon_{2z}|}  \,.
\end{equation}
The loss-induced angle switching observed in Fig.~\ref{Fig2DTang2} can be explained from Eq.~(\ref{tantheS}).  For the anisotropic material with $\epsilon_{2x}\ne\epsilon_{2z}$ and finite $\epsilon_{2x}$, if $\epsilon_{2z}^i=0$, when $\epsilon_{2z}^r\rightarrow0$, $\epsilon_{2z}^r/|\epsilon_{2z}|\rightarrow1$ and $a\rightarrow0$, thus $\theta_S\rightarrow90^\circ$.  If $\epsilon_{2z}^i\ne0$, when $\epsilon_{2z}^r\rightarrow0$, $\epsilon_{2z}^r/|\epsilon_{2z}|\rightarrow0$ and $a$ is finite, thus $\theta_S\rightarrow0^\circ$.  On the other hand, if $\epsilon_{2z}$ is finite, when $\epsilon_{2x}\rightarrow0$, $a\rightarrow\sqrt{\epsilon_{2x}}$, thus $\theta_S\rightarrow0^\circ$.  For the isotropic case, let $\epsilon_{2x}=\epsilon_{2z}\equiv\epsilon_2^r+i\epsilon_2^i$.  If $\epsilon_2^i=0$, when $\epsilon_2^r\rightarrow0$, $\epsilon_{2z}^r/|\epsilon_{2z}|\rightarrow1$ and $a\rightarrow(\epsilon_2^r)^{3/2}$, thus $\theta_S\rightarrow90^\circ$.  If $\epsilon_2^i\ne0$, when $\epsilon_2^r\rightarrow0$, $\epsilon_{2z}^r/|\epsilon_{2z}|\rightarrow0$ and $a$ is finite, therefore $\theta_S\rightarrow0^\circ$.  To validate Eq.~(\ref{tantheS}), in Fig.~\ref{Fig2DTang2} the TAs calculated from Eq.~(\ref{tantheS}) (green-circles) are compared to those computed numerically (blue-solid), showing a perfect agreement.

\begin{figure}[htb]
\centering\includegraphics[height=2.5in, width=3in]{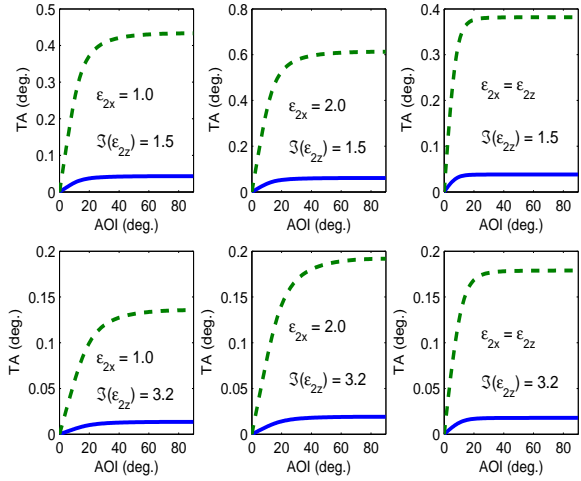}
\caption{(Color online) Transmission angle versus AOI when the $\Re(\epsilon_{2z})=0.001$ (blue-solid) and $\Re(\epsilon_{2z})=0.01$ (green-dashed).  Top panels: $\Im(\epsilon_{2z})=1.5$.  Bottom panels: $\Im(\epsilon_{2z})=3.2$.
Left and middle panels: anisotropic ENZ material with $\epsilon_{2x}=1.0$ (left panels) and $\epsilon_{2x}=2.0$ (middle panels).  Right panels: isotropic ENZ material.  The permittivity of the input medium $\epsilon_1=36$.}
\label{Fig3DTang3}
\end{figure}

To validate the loss-induced switching behavior is a robust feature, in Fig.~\ref{Fig3DTang3} the transmission angle versus AOI is plotted for the different real parts of $\epsilon_{2z}$ and $\epsilon_{2x}$ and the material loss.  In essence, the transmission angle decreases with increasing the loss $\Im(\epsilon_{2z})$ and decreasing the $\Re(\epsilon_{2z})$.
When $\Re(\epsilon_{2z})\rightarrow0$, the angular width of the transmission can be estimated from
\begin{equation}
\label{Dthe}
\Delta\theta_S\approx \left\{ \begin{matrix}
\dfrac{\sqrt{2}\,\,|\epsilon_x|\,\epsilon_z^r} {|\epsilon_z|^{3/2} \sqrt{|\epsilon_x| + \epsilon_x^i + \eta\,\epsilon_x^r}}  \,, &  \mbox{if } \eta\leq1  \\  \\
\dfrac{\sqrt{2}\,\,|\epsilon_x|\,\epsilon_z^r} {|\epsilon_z|^{3/2} \sqrt{\epsilon_x^i + \eta\bigl(|\epsilon_x|+\epsilon_x^r\bigr)}}  \,, &  \mbox{if } \eta\geq1 
\end{matrix} \right.  \,,
\end{equation}
where $\eta\equiv\dfrac{|\epsilon_z|\mu_y}{\epsilon_1\mu_1}$, and the subscript $2$ in the $\epsilon_x$, $\epsilon_z$, and $\mu_y$ was omitted in above equation.
\begin{figure}[htb]
\centering\includegraphics[height=2.5in, width=3in]{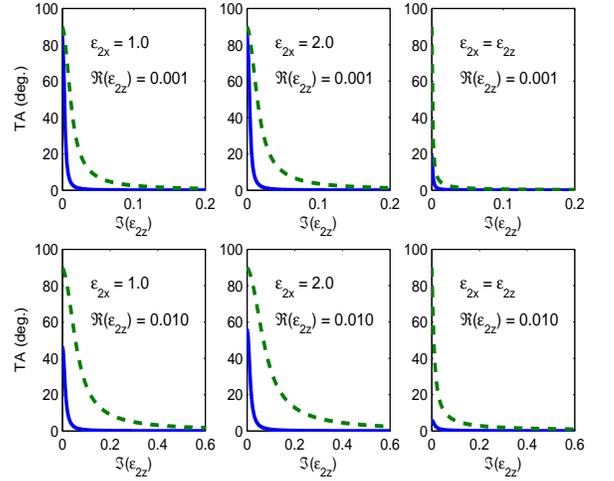}
\caption{(Color online) Transmission angle versus material loss $\Im(\epsilon_{2z})$ computed for AOI $=0.1^\circ$ (blue-solid) and AOI $=89^\circ$ (green-dashed).  Top panels: $\Re(\epsilon_{2z})=0.001$.  Bottom panels: $\Re(\epsilon_{2z})=0.01$.  Left and middle panels: anisotropic ENZ material with $\epsilon_{2x}=1.0$ (left panels) and $\epsilon_{2x}=2.0$ (middle panels).  Right panels: isotropic ENZ material.  The permittivity of the input medium $\epsilon_1=36$.  TA quickly converges to zero in all the scenarios.}
\label{Fig4DTang6}
\end{figure}
Figure~\ref{Fig4DTang6} demonstrates how rapidly the transmission angle converges to zero as the loss $\Im(\epsilon_{2z})$ increases for the different values of $\Re(\epsilon_{2z})$ and $\epsilon_{2x}$.  
The blue-solid curves represent the transmission angles calculated for the near-zero angle of incidence, while the green-dashed curves for the grazing angle of incidence.  
The difference between the green-dashed and blue-solid curves corresponds to the angular width of the transmission.  
The angular width in the isotropic ENZ medium (right panels) is usually smaller than that in the anisotropic medium (left and middle panels).  This is implicated in Eq.~(\ref{Dthe}) as well.
It is well-known that loss is inextricable to metamaterial.  Many fascinating effects diminish as the result of high loss \cite{Dimmock,Nieto}.  
However, for the effect demonstrated here, the material loss plays a positive role, resulting in the omnidirectional bending of light towards the normal upon entering the ENZ medium.

This phenomenon may have many applications, such as directive antennas.  Instead of radiation applications, we will explore this phenomenon from a receiving perspective, i.e., redirect the incoming EM power from different directions to the direction of the receivers to enhance the acquisition power, as illustrated in Fig.~\ref{Fig1DTdraw}b.  To increase the coupling, a matching coating can be deposited on the surface of the ENZ medium such that the effective impedance of the overall structure is matched to the free-space impedance.  
For simplicity, here a dielectric-metal-dielectric thin film is superimposed on the ENZ material.  This sandwich structure possesses coupled surface plasmon modes due to closely spaced two dielectric-metal interfaces.  By exciting the plasmonic resonances of the structure, the transmission can be enhanced.  The resonant frequency of the transmission can be tuned by changing the thickness of the layers.  In our simulation, the materials of the dielectric and metallic layers are, respectively, amorphous polycarbonate (APC) and silver (Ag).  The refractive index of the APC is given by \cite{Roberts}
\begin{equation}
n_p =  1.5567 + \frac{8.0797\times10^{-3}}{\lambda^2} + \frac{3.5971\times10^{-4}}{\lambda^4}  \,,
\end{equation}
where $\lambda$ is the wavelength in $\mu m$.  The loss of the APC is very small in the wavelength range of the simulation, and thus is neglected \cite{Roberts}.  The absorption of Ag is included through the complex permittivity given from Palik \cite{Palik}. 
\begin{figure}[htb]
\centering\includegraphics[width=.4\textwidth]{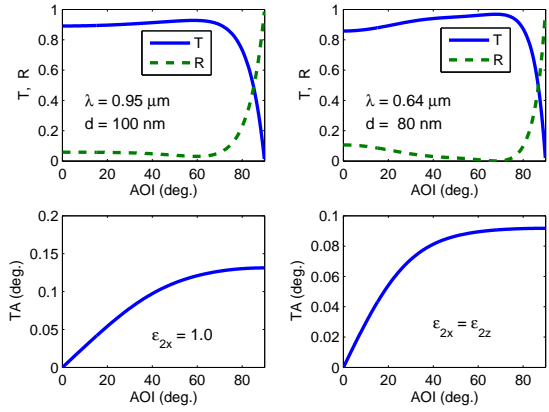}
\caption{(Color online) Top panels: Transmittance (blue-solid) and reflectance (green-dashed) of the APC-Ag-APC thin film versus AOI when the medium after the film is the anisotropic ENZ material with $\epsilon_{2x}=1$ (left panels) or the isotropic ENZ material (right panels).  $\epsilon_{2z}=0.001+0.6i$ for both cases.  Bottom panels: transmission angle (corresponding to the top panels) versus AOI.  The thickness of the APC $d=100\,$nm (left panels) and $d=80\,$nm (right panels).  The thickness of Ag is 10 nm for both cases.}
\label{Fig5DTpow5}
\end{figure}
\begin{figure}[htb]
\centering\includegraphics[height=1.6in, width=3in]{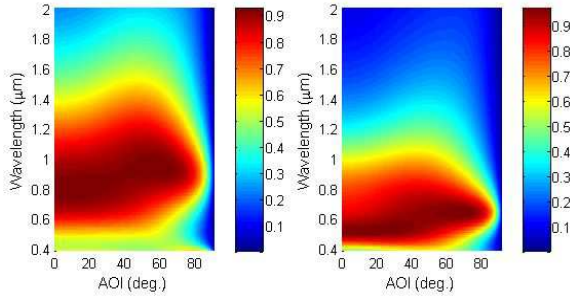}
\caption{(Color online) Transmittance of the APC-Ag-APC thin film versus AOI and wavelength when the medium after the film is the anisotropic (left panel) or the isotropic (right panel) ENZ material.  Color-bars represent the magnitude of the transmittance.  Simulation parameters are the same as those in Fig.~\ref{Fig5DTpow5}.}
\label{Fig6DTpow6}
\end{figure}

Shown in Fig.~\ref{Fig5DTpow5} are the transmission and reflection (top panels) of a plane wave incident from air to the APC-Ag-APC structure, along with the corresponding transmission angle (bottom panels).  At the resonance, the thickness of the APC $d=100\,$nm with the resonant wavelength $\lambda=0.95\,\mu$m for the anisotropic ENZ medium (left panels); and the $d=80\,$nm with the $\lambda=0.64\,\mu$m for the isotropic ENZ medium (right panels).  About $90\%$ transmittance are achieved for a wide range of the incidence angle up to $70^\circ$ (see top panels) with nearly-collimated transmission in the normal direction (see bottom panels).  Transmittance of the APC-Ag-APC thin film as a function of AOI and wavelength is presented in Fig.~\ref{Fig6DTpow6} when the medium at the back of the film is the anisotropic (left panel) or the isotropic (right panel) ENZ material.  In both cases, wide-angle $90\%$ transmittance are observed.
It is worth mentioning that the loss of the ENZ medium does not affect the transmittance of the APC-Ag-APC structure since the transmitted power is computed right after the thin film, i.e., before traveling through the ENZ medium.  If the receptors are embedded very close to the back of the film, the propagation loss in the ENZ medium can be minimized.  However, the loss of the ENZ material plays an important role on controlling the direction of the transmission, no matter where the receptors are located.

In conclusions, we have demonstrated omnidirectionally transmitting the electromagnetic power to one direction in the ENZ materials.  This phenomenon is realized based on two mechanisms.  One is the loss-assistant bending of the EM power to the normal for all angles of incidence.  The other is the enhanced transmission through structural resonances.  This phenomenon may have applications in communications, directive antennas, as well as detectors and sensors to increase acceptance angle and redirect electromagnetic power without using optical lenses and mechanical gimbals.  The concept of employing metamaterial loss to control the direction of the transmission brings a positive perspective for material loss and may open up a new avenue for metamaterial designs and applications.

The author gratefully acknowledge the sponsorship of ONR's N-STAR and NAVAIR's ILIR programs.

\end{document}